\documentclass[prl,twocolumn,showpacs]{revtex4-1}

\usepackage{graphicx}
\usepackage{amsmath}

\newcommand{\DL}{D_{\rm L}}

\newcommand{\DT}{D_{\rm T}}
\newcommand{\eg}{{e.g., }}
\newcommand{\etaf}{\eta_{\rm f}}
\newcommand{\etam}{\eta_{\rm m}}
\newcommand{\Geff}{G^{\rm eff}}
\newcommand{\Gf}{G^{\rm free}}

\newcommand{\ie}{{i.e., }}
\newcommand{\kT}{k_{\rm B}T}

\newcommand{\phimax}{\phi_{\rm max}}
\newcommand{\tG}{{\bf G}}
\newcommand{\tGeff}{{\bf G}^{\rm eff}}
\newcommand{\tGf}{{\bf G}^{\rm free}}
\newcommand{\tGs}{{\bf G}^{\rm sup}}

\newcommand{\tlGeff}{\tilde{G}^{\rm eff}}
\newcommand{\tlGf}{\tilde{G}^{\rm free}}
\newcommand{\tlGs}{\tilde{G}^{\rm sup}}
\newcommand{\ttG}{\tilde{\bf G}}

\newcommand{\vecF}{{\bf F}}
\newcommand{\vecq}{{\bf q}}
\newcommand{\vecr}{{\bf r}}
\newcommand{\vecv}{{\bf v}}
\newcommand{\xhat}{\hat{\bf x}}

\begin{document}

\title{In-Plane Dynamics of Membranes with Immobile Inclusions}

\author{Naomi Oppenheimer}
\email{naomiopp@tau.ac.il}
\affiliation{Raymond \& Beverly Sackler School of Chemistry, Tel Aviv
University, Tel Aviv 69978, Israel}

\author{Haim Diamant} 
\email{hdiamant@tau.ac.il} 
\affiliation{Raymond \& Beverly Sackler School of Chemistry, Tel Aviv
University, Tel Aviv 69978, Israel}

\date{\today}

\begin{abstract}
  Cell membranes are anchored to the cytoskeleton via immobile
  inclusions.  We investigate the effect of such anchors on the
  in-plane dynamics of a fluid membrane and mobile inclusions
  (proteins) embedded in it.  The immobile particles lead to a
  decreased diffusion coefficient of mobile ones and suppress the
  correlated diffusion of particle pairs.  Due to the long-range,
  quasi-two-dimensional nature of membrane flows, these effects become
  significant at a low area fraction (below one percent) of immobile
  inclusions.
\end{abstract}


\maketitle

Biomembranes are key structural and functional ingredients of any
living cell. They are based on a self-assembled fluid bilayer of
amphiphilic molecules (lipids), containing a high concentration of
embedded proteins that perform vital biological tasks \cite{biology}.
From a physical viewpoint, biomembranes represent an interesting state
of matter, whose properties are intermediate between two-dimensional
(2D) and three-dimensional (3D), and between fluid and solid. In the
past four decades the structural properties of membranes and the
dynamics of their out-of-plane fluctuations have been well
characterized \cite{SafranBook,Lipowsky}. The in-plane fluid dynamics,
which is crucial, \eg for the diffusion of membrane proteins, was
first addressed by Saffman and Delbr\"uck (SD) \cite{SD}. Their model
considered the motion of a single inclusion within a viscous slab
surrounded by an infinite fluid. While certain experiments were
consistent with the SD predictions \cite{Cicuta2007,Ramadurai2009},
other works pointed to discrepancies \cite{Gambin2006,Naji2007}.  The
SD model was extended in various directions, \eg to treat proteins of
larger size \cite{Hughes1980} and number \cite{ourBJ}, viscoelastic
effects \cite{Levine2002,Camley2011}, in-plane concentration
fluctuations and phase separation
\cite{Inaura2008,Ramachandran2010,Camley2010,Fan2010}, and the effect
of an adjacent rigid surface
\cite{StoneAjdari,ourPRE,Ramachandran2011}.
Contrary to these models, most biomembranes are neither freely
suspended in an infinite fluid nor attached to a rigid wall. Rather,
they are anchored via proteins or protein-associated domains to a
dilute, soft network of filaments (cortical actin in animal cells or
cortical microtubules in plant cells) \cite{biology}. Our aim is to
explore how the membranal in-plane response, and the consequent
dynamics of mobile inclusions, are affected by such immobile anchors.

Models of membrane dynamics are divided into those that conserve
momentum in 3D
\cite{SD,Hughes1980,Levine2002,ourBJ,Inaura2008,Camley2010,Ramachandran2010,Fan2010},
as applicable to freely suspended membranes, 
and those that do not
\cite{EvansSackmann,Suzuki,Nelson,StoneAjdari,ourPRE,Ramachandran2011},
as appropriate for substrate-supported membranes. This essential
difference is reflected in the respective velocity Green's functions,
$\tGf(\vecr)$ and $\tGs(\vecr)$\,---\,tensors that give the flow
velocity of the membranal fluid at position $\vecr$, $\vecv(\vecr)$,
in response to a localized in-plane force, $\vecF$, applied at the
origin, $v_i(\vecr)=G^{\rm free/sup}_{ij}(\vecr)F_j$ (with $i,j=x,y$
and summation over repeated indices). In Fourier space,
$\ttG(\vecq)=\int d^2r e^{-i\vecq\cdot\vecr}\tG(\vecr)$, these are
\cite{ourBJ,ourPRE}
\begin{eqnarray}
  \tlGf_{ij} = \frac{1}{\etam q(q+\kappa)} \left(\delta_{ij} 
  - \frac{q_iq_j}{q^2} \right)&&,
\label{tGf}\\
  \tlGs_{ij} = \frac{1}{\etam (q^2+\alpha^2)} \left(\delta_{ij} 
  - \frac{q_iq_j}{q^2} \right)&&,
\label{tGs}
\end{eqnarray}
where we have assumed a flat, incompressible membrane with 2D
viscosity $\etam$. In Eq.\ (\ref{tGf})
$\kappa^{-1}$ is the SD length, the characteristic distance beyond
which the free membrane exchanges momentum with the surrounding fluid
\cite{SD}.  In the SD model $\kappa^{-1}=\etam/(2\etaf)$, where
$\etaf$ is the viscosity of the outer fluid. Typical values for the
membrane 2D viscosity are $\etam\sim 10^{-10}$--$10^{-9}$
Pa$\cdot$s$\cdot$m, yielding $\kappa^{-1}\sim 0.1$--$1$ $\mu$m. In
Eq.\ (\ref{tGs}) $\alpha^{-1}$ is the distance beyond which the
supported membrane loses momentum to the substrate.  (For a membrane
lying a distance $h$ from a rigid wall, $\alpha=[\kappa/(2h)]^{1/2}$
\cite{ourPRE}.)  Despite the appearance of the decay parameters
$\kappa$ and $\alpha$, Eqs.\ (\ref{tGf}) and (\ref{tGs}) describe
long-ranged velocity responses\,---\,$\tGf$ decays as $1/r$ and $\tGs$
as $1/r^2$ (due to the conservation of 3D momentum in the former, and
2D membrane mass in the latter \cite{JPSJ09}).

Biomembranes contain anchoring inclusions separated by typical
distances of $30$--$80$ nm \cite{biology}, \ie covering an area
fraction $\phi\sim 10^{-3}$--$10^{-2}$. These immobile inclusions
break the membrane's translational symmetry and absorb momentum;
hence, the large-distance response should be similar to $\tGs$. From
another perspective, we expect the immobile obstacles to serve as a
porous matrix within which the lipids flow.  Indeed, Eq.\ (\ref{tGs})
is analogous to the response of a fluid embedded in a porous medium
\cite{JPSJ09}.  We establish below that this intuitive picture is
correct and derive the momentum decay length as a function of the area
fraction $\phi$ of immobile inclusions. As this length decreases from
arbitrarily large to smaller values upon increasing $\phi$, the
membrane crosses over from a dominantly ``free'' behavior to a
``supported'' (or ``porous'') one.

We first note the quasi-2D nature of the system.  Over sufficiently
small distances a membrane behaves as a momentum-conserving 2D fluid,
whose velocity Green's function is given by \cite{Pozrikidis}
\begin{equation}
  G^{\rm 2D}_{ij}(\vecr) = \frac{1}{4\pi\etam} \left\{ \left[
  \ln\left(\frac{2}{\beta r}\right)-\gamma-\frac{1}{2} \right] \delta_{ij}
  + \frac{r_ir_j}{r^2} \right\}, 
\label{G2D}
\end{equation}
where $\beta^{-1}$ is a cutoff length regularizing the divergent 2D
behavior and $\gamma\simeq0.577$ is Euler's constant. Both Eqs.\ 
(\ref{tGf}) and (\ref{tGs}), upon inversion to real space and taking
the limit of small distances, coincide with $G^{\rm 2D}$, with
$\beta=\kappa$ or $\beta=\alpha$, correspondingly.  In the current
problem the immobile inclusions produce an effective cutoff for the
membrane's 2D behavior, to be derived below. With increasing $\phi$,
$\beta^{-1}$ decreases from $\kappa^{-1}$ down to a microscopic
length.

Let us consider a bare free membrane and ask how its velocity
response, Eq.\ (\ref{tGf}), is modified by the presence of immobile
cylindrical inclusions \cite{ft_shape}. The inclusion radius, $a$, is
taken as the smallest length in the problem, and the calculation is
restricted to the leading terms in $\kappa a$ and $a/r$. We begin by
applying a localized force $\vecF$ at the origin. The resulting flow
velocity of the bare membrane is
$v_i^{(0)}(\vecr)=\Gf_{ij}(\vecr)F_j$.  Next we place an immobile
inclusion at position $\vecr'$. Due to the flow $\vecv^{(0)}$, the
inclusion exerts an additional force on the membrane,
$\vecF^{(1)}\simeq -\Gamma\vecv^{(0)}(\vecr')$, where
\begin{equation}
 \Gamma=4\pi\etam/\{\ln[2/(\beta a)]-\gamma\}
\label{Gamma}
\end{equation}
is the friction coefficient of the cylindrical \cite{ft_shape}
inclusion \cite{SD,EvansSackmann,ourBJ}, $\beta$ being an effective 2D
cutoff.  We neglect moments of the force distribution higher than this
monopole.
The inclusion-induced force changes the flow velocity by
$v^{(1)}_i(\vecr)=\Gf_{ij}(\vecr-\vecr')F^{(1)}_j(\vecr')$. Thus, a
single immobile inclusion reflects the original force as
$v^{(1)}_i(\vecr)=-\Gamma\Gf_{ij}(\vecr-\vecr')\Gf_{jk}(\vecr')F_k$.
Now, consider many randomly distributed immobile inclusions, covering
an area fraction $\phi$ of the membrane. To first order in $\phi$,
their effect on the membrane's flow is $\langle
v^{(1)}_i(\vecr)\rangle = G^{(1)}_{ij}(\vecr)F_j$, with
$G^{(1)}_{ij}(\vecr) = -\Gamma[\phi/(\pi a^2)]\int
d^2r'\Gf_{ik}(\vecr-\vecr')\Gf_{kj}(\vecr')$.

At a higher value of $\phi$ many-body terms set in\,---\,the flow
reflected from one inclusion is reflected again from another, and so
on. We have calculated all orders of these monopolar hydrodynamic
terms while continuing to assume a uniform static distribution of
inclusions \cite{suppl}. 
This yields $v_i(\vecr)=\Geff_{ij}(\vecr)F_j$, with a cumbersome
expression for $\tGeff(\vecr)$ \cite{suppl}. In Fourier space,
\begin{eqnarray}
  \tlGeff_{ij}(\vecq) &=& \frac{1}{\etam[q(q+\kappa)+\lambda^{-2}]} \left(\delta_{ij} 
  - \frac{q_iq_j}{q^2} \right), 
\label{Geff}\\
  \lambda &=& a [\Gamma\phi/(\pi\etam)]^{-1/2}.
\label{lambda}
\end{eqnarray}
In the limit $r\rightarrow 0$, $\tGeff$ has the expected form of
$\tG^{\rm 2D}$ [Eq.\ (\ref{G2D})], with a cutoff $\beta$ that
satisfies
\begin{equation}
  \ln(\beta\lambda) = \frac{\tanh^{-1}[f(\kappa\lambda/2)]}{f(\kappa\lambda/2)},\ \ 
  f(x) = \frac{\sqrt{x^2-1}}{x}.
\label{SC}
\end{equation}

\begin{figure}[tbh]
\vspace{0.5cm}
\centerline{
\resizebox{0.4\textwidth}{!}{\includegraphics{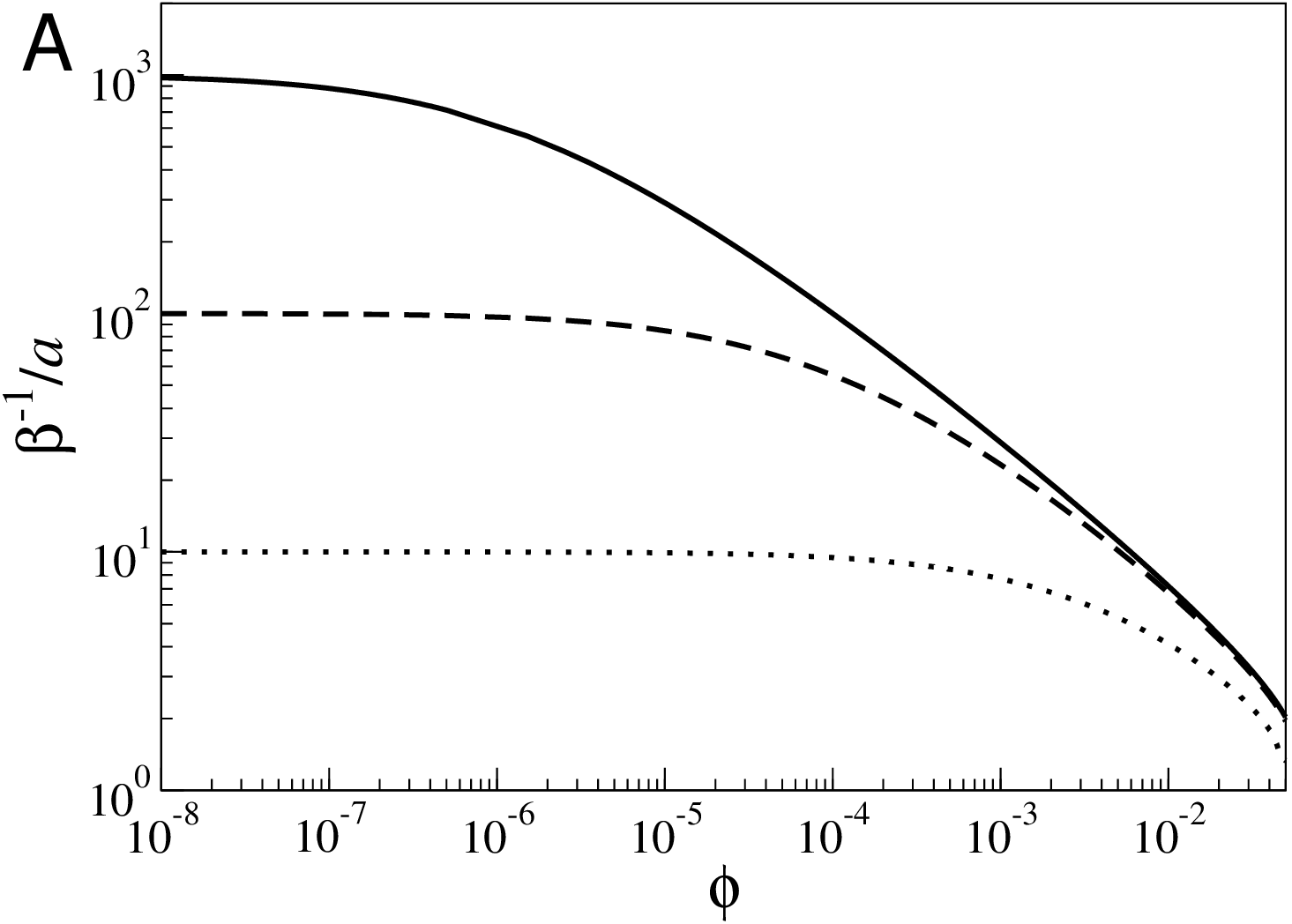}}}
\vspace{0.9cm}
\centerline{
\resizebox{0.4\textwidth}{!}{\includegraphics{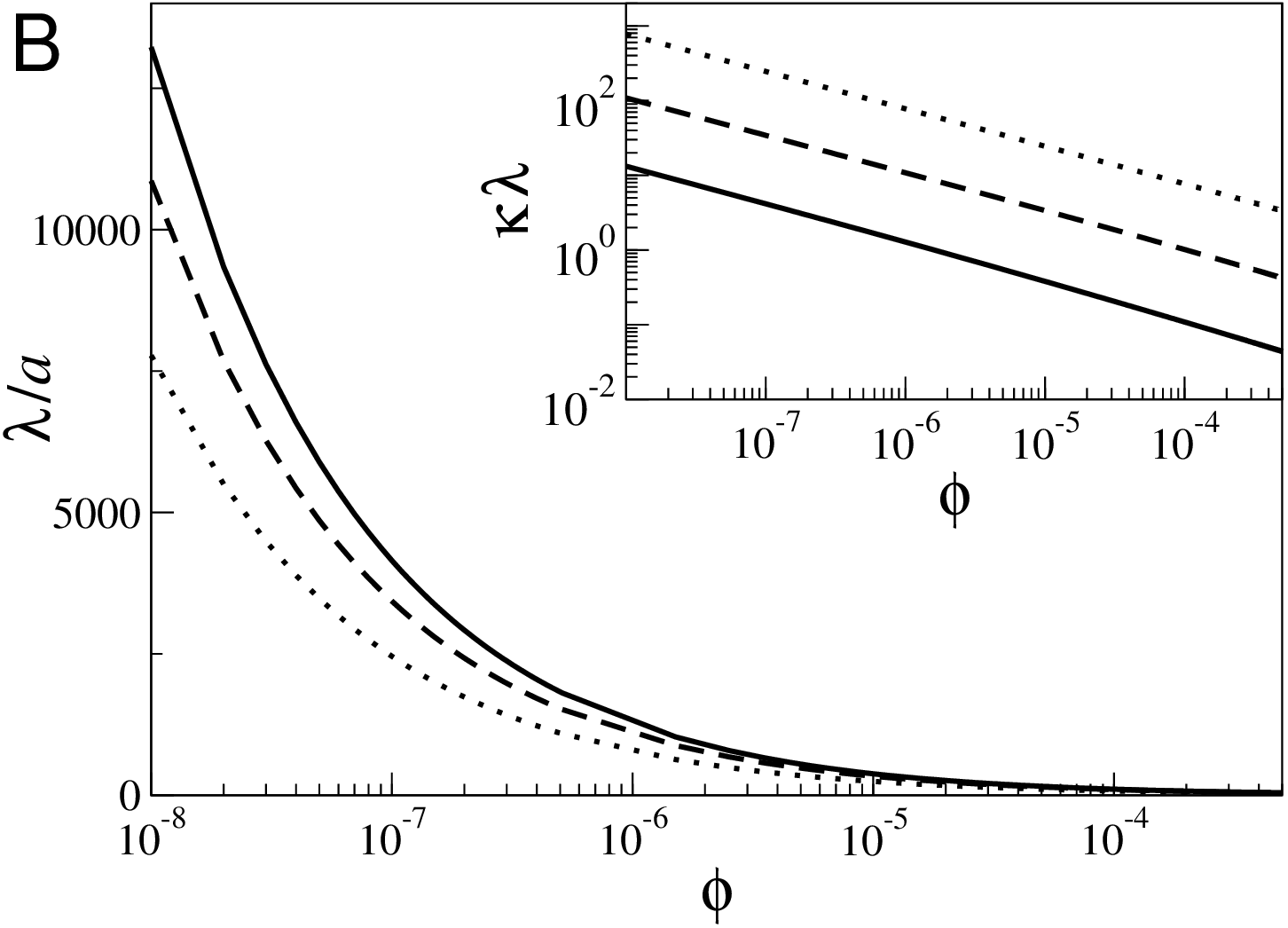}}}
\caption{Two-dimensional cutoff length (A) and momentum decay length
  (B) as a function of area fraction of immobile inclusions. Solid,
  dashed, and dotted curves correspond to $\kappa a=10^{-3}$,
  $10^{-2}$, and $10^{-1}$, respectively. The lengths are scaled by
  the inclusion size $a$. The inset of panel B shows $\lambda$, scaled
  by the SD length $\kappa^{-1}$, on a log-log plot, demonstrating a
  roughly $\phi^{-1/2}$ decay.}
\label{fig_beta}
\end{figure}

Equations (\ref{Gamma}), (\ref{lambda}), and (\ref{SC}) provide a
self-consistent scheme for obtaining $\beta a$ and $\lambda/a$ as
functions of $\phi$ and $\kappa a$. The decrease of $\beta^{-1}$ and
$\lambda$ with increasing $\phi$ is shown in Fig.\ 
\ref{fig_beta}.  In the limit $\phi\rightarrow 0$ we have
$\lambda\rightarrow\infty$ [Eq.\ (\ref{lambda})], and Eq.\ (\ref{SC})
then gives $\beta\rightarrow\kappa$ (see Fig.\ 
\ref{fig_beta}A)\,---\,\ie the cutoff of the 2D behavior is that of a
free membrane, as expected. As $\phi$ increases, $\lambda$ decreases
roughly as $\phi^{-1/2}$ (Fig.\ \ref{fig_beta}B inset), while
$\beta^{-1}$ decreases initially as $\phi\ln\phi$, and subsequently
more sharply (Fig.\ \ref{fig_beta}A).  Both lengths reach values
comparable to the inclusion size $a$, whereupon the theory breaks
down. The breakdown is marked by a loss of solutions for the
self-consistent scheme, occurring at a fixed (up to corrections of
order $\kappa a$), small area fraction, $\phimax\simeq 0.058$.  This
analysis reveals two distinct regimes (cf.\ Fig.\ \ref{fig_beta}): a
low-concentration regime, $\phi\ll(\kappa a)^2|\ln(\kappa a)|$, in
which $\lambda\gg\beta^{-1}\simeq\kappa^{-1}$, and a
higher-concentration regime, $(\kappa a)^2|\ln(\kappa a)| \ll \phi <
\phimax$, where $\lambda\sim\beta^{-1}\ll\kappa^{-1}$.

\begin{figure}[tbh]
\vspace{0.5cm}
\centerline{
\resizebox{0.4\textwidth}{!}{\includegraphics{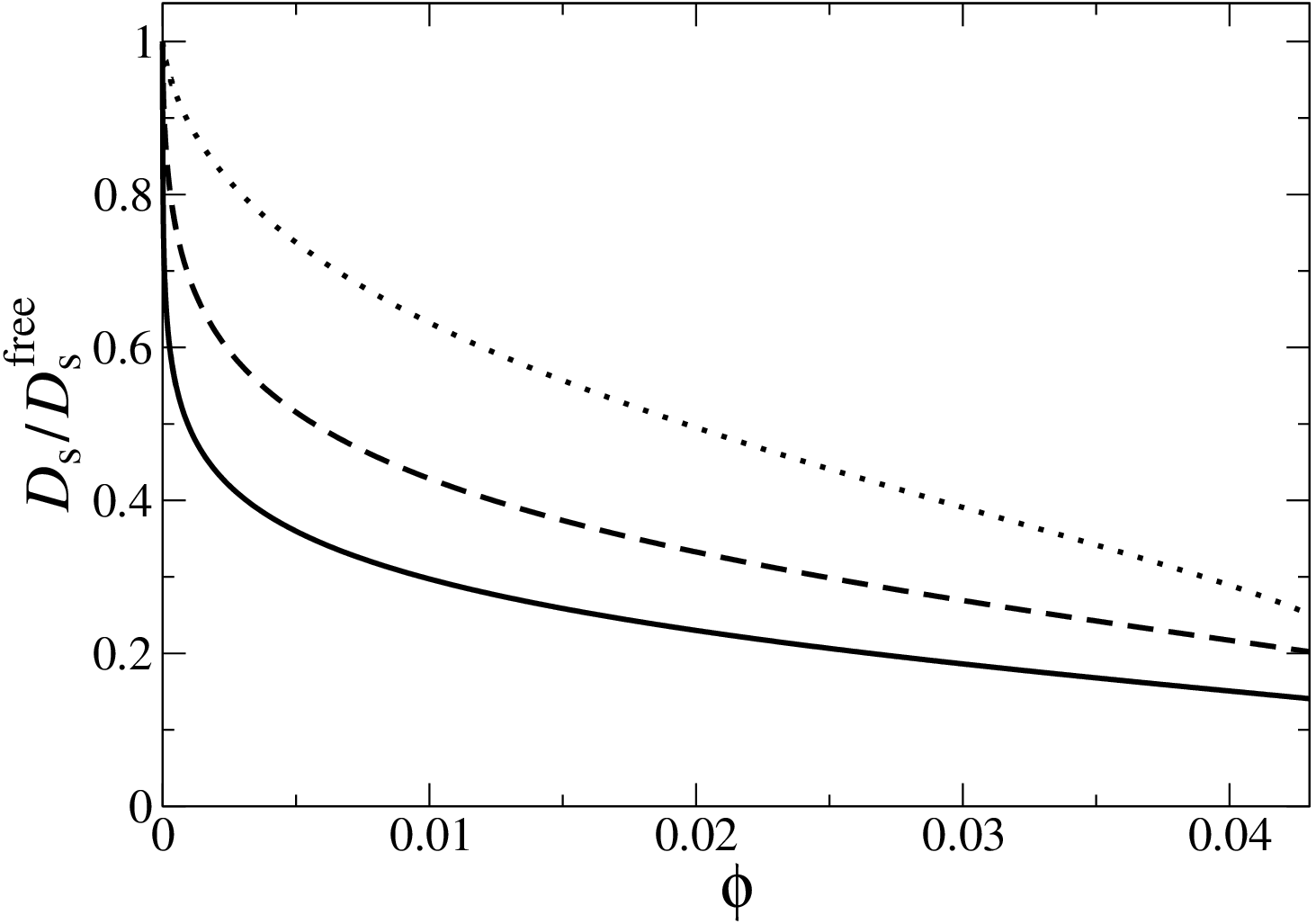}}}
\caption{Self-diffusion coefficient of a mobile inclusion as a
  function of area fraction of immobile inclusions. The coefficient is
  scaled by its value in a free membrane ($\phi=0$). The radii of the
  two types of inclusion are taken as equal, $b=a$. Solid, dashed, and
  dotted curves correspond to $\kappa a=10^{-3}$, $10^{-2}$, and
  $10^{-1}$, respectively.}
\label{fig_Ds}
\end{figure}

We can readily examine the effect of $\phi$ on the self-diffusion
coefficient of a mobile inclusion of radius $b$. From Eq.\ 
(\ref{Gamma}) and Einstein's relation, $D_{\rm s}=\kT \{\ln[2/(\beta
b)]-\gamma\}/(4\pi\etam)$, where $\kT$ is the thermal energy. Because
of the increase of $\beta$ with $\phi$, $D_{\rm s}$ decreases from its
free value $D_{\rm s}^{\rm free}$ as given by SD \cite{SD} (the
expression above with $\beta=\kappa$, yielding typical values of
$D_{\rm s}^{\rm free}\sim 1$--$10$ $\mu$m$^2$/s), to significantly
lower values (Fig.\ \ref{fig_Ds}).

Next we address the coupling diffusion coefficients of a pair of
mobile inclusions separated by a distance $r$. The longitudinal
diffusion coefficient, $\DL(r)=\langle \Delta x_1\Delta
x_2\rangle/(2t)$, and the transverse one, $\DT(r)=\langle \Delta
y_1\Delta y_2\rangle/(2t)$, characterize the correlated diffusion of
the pair, respectively, along and perpendicular to their connecting
line. Here $\Delta x_i$ and $\Delta y_i$ are the displacements of
particle $i$ along and perpendicular to the connecting line during
time $t$.
In the limit $r\gg b$, the pair diffusion coefficients are directly
obtained as $\DL(r)\simeq\kT\Geff_{xx}(r\xhat)$ and
$\DT(r)\simeq\kT\Geff_{yy}(r\xhat)$, where $\xhat$ is a unit vector
along the connecting line. As is evident from Eq.\ (\ref{Geff}), the
coupling diffusion coefficients in this limit depend only on $\lambda$
and $\kappa$; they are independent of the size of mobile inclusions
and depend on the size of the immobile ones only indirectly, through
$\lambda$.

\begin{figure}[tbh]
\vspace{0.6cm}
\centerline{
\resizebox{0.4\textwidth}{!}{\includegraphics{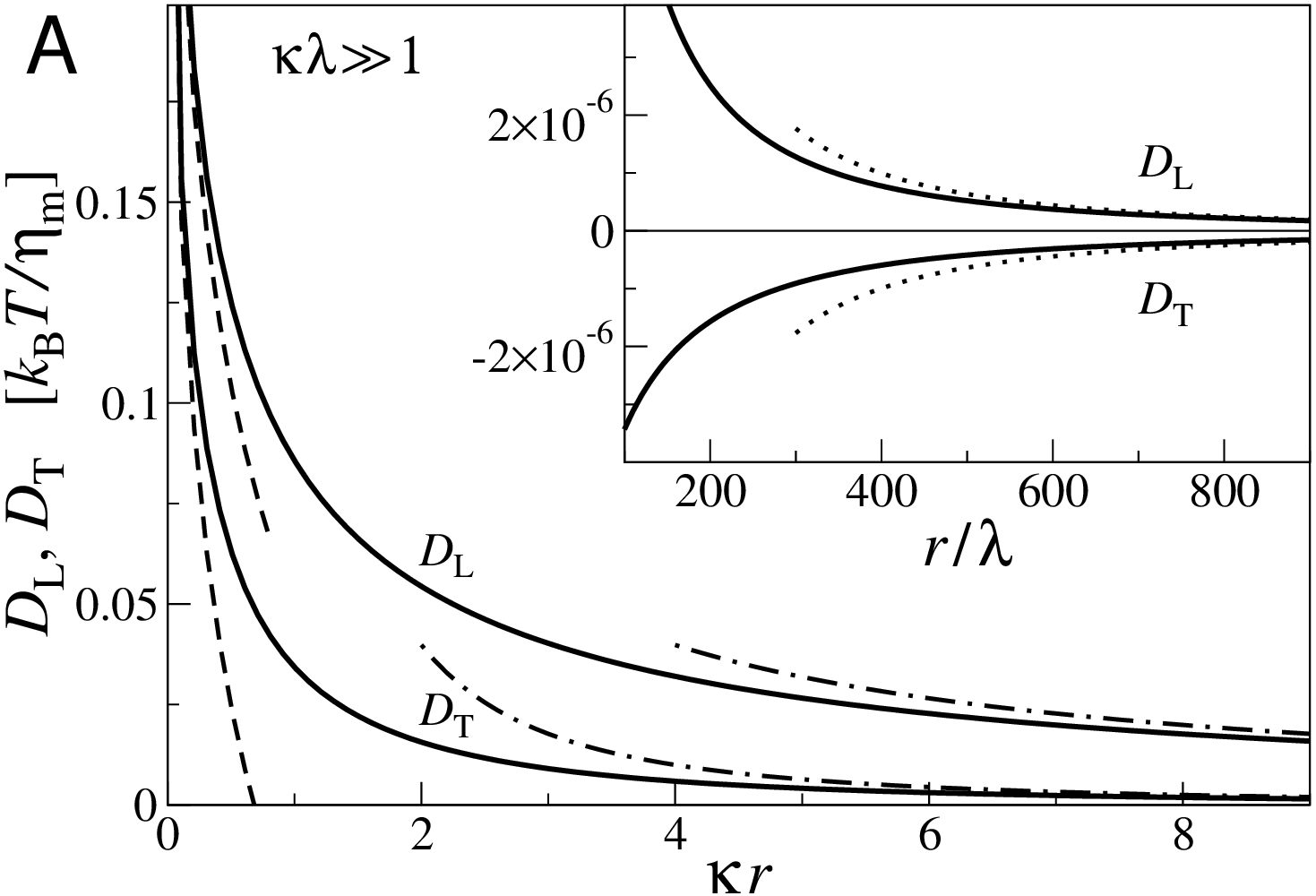}}}
\vspace{0.9cm}
\centerline{
\resizebox{0.4\textwidth}{!}{\includegraphics{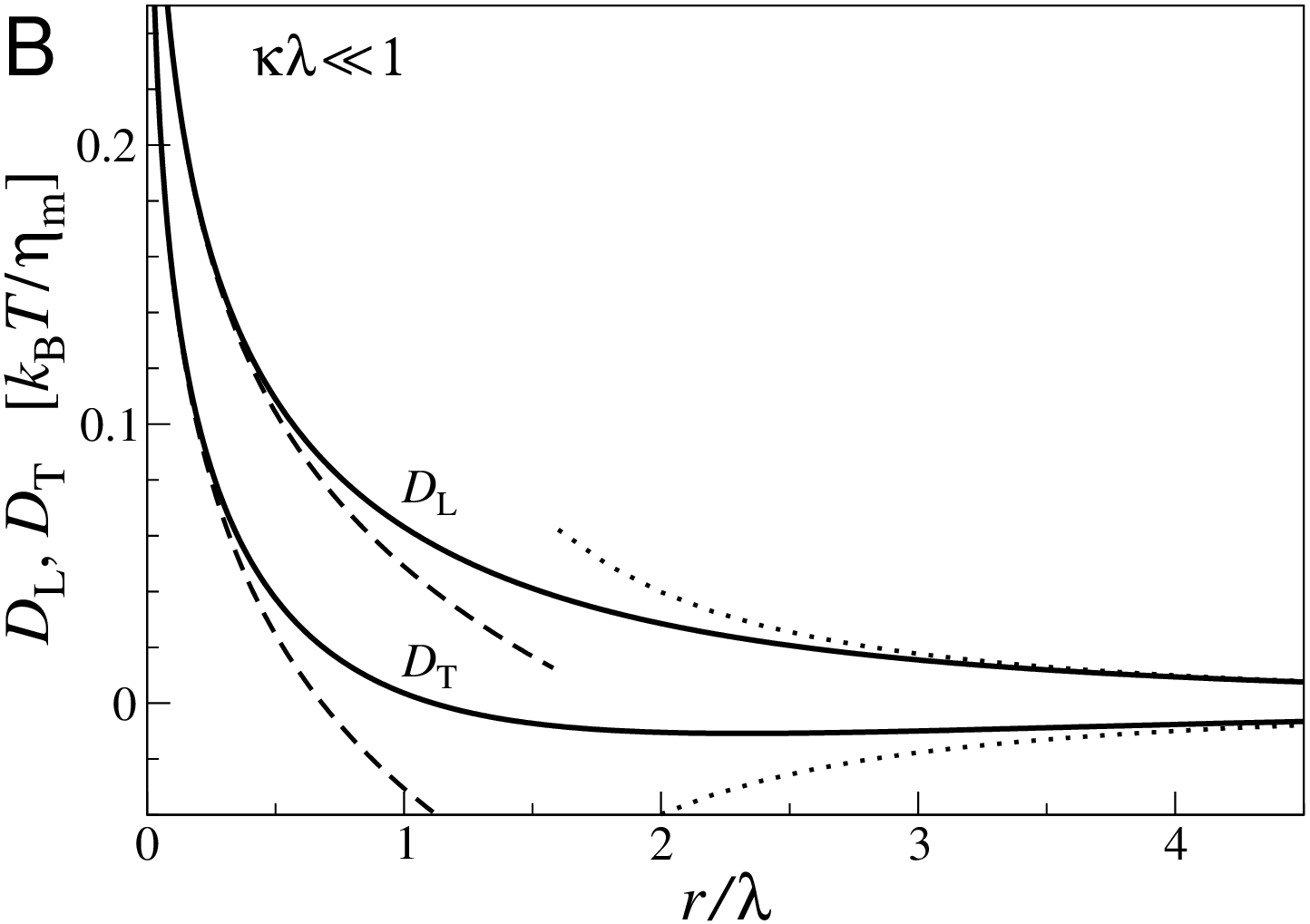}}}
\caption{Coupling diffusion coefficients of a pair of mobile
  inclusions as a function of mutual distance. Panels A and B present,
  respectively, results for low concentration ($\kappa\lambda=100$)
  and high concentration ($\kappa\lambda=0.01$) of immobile
  inclusions. Results obtained from the full Green's function (solid
  lines) are shown together with their various asymptotes (see
  text). The coupling diffusion coefficients are scaled by
  $\kT/\etam$, and the distance by either the Saffman-Delbr\"uck
  length $\kappa^{-1}$ or the momentum decay length $\lambda$.}
\label{fig_DLT}
\end{figure}

The results obtained for $\DL(r)$ and $\DT(r)$ using the full
expression for $\tGeff(\vecr)$ \cite{suppl} are shown in Fig.\ 
\ref{fig_DLT} (solid lines). They include several simpler asymptotic
regions.  The low-concentration regime ($\kappa\lambda\gg 1$, Fig.\ 
\ref{fig_DLT}A) includes three such regions. At short distances
($r\ll\kappa^{-1}$) the behavior is 2D-like, resulting in
$\DL,\DT\sim|\ln(\beta r)|$ with $\beta\simeq\kappa$ (dashed lines). At
intermediate distances ($\kappa^{-1}\ll r\ll\lambda$) the coupling
becomes 3D-like, with $\DL\sim 1/r$ and $\DT\sim 1/r^2$ (dash-dotted
lines). In these two regions the coefficients coincide with those in a
free membrane, as obtained from Eq.\ (\ref{tGf}) \cite{ourBJ}. At
sufficiently large interparticle distances ($r\gg\lambda$) the
coupling becomes sensitive to the immobile inclusions, with
$\DL,\DT\sim\pm 1/r^2$ (dotted curves in the inset). These coefficients
coincide with those in a supported membrane, as obtained from Eq.\ 
(\ref{tGs}) with $\alpha=\lambda^{-1}$ \cite{ourPRE}. The
higher-concentration regime ($\kappa\lambda\ll 1$, Fig.\ 
\ref{fig_DLT}B) contains two asymptotic regions. At short distances
($r\ll\lambda$) the behavior is again 2D-like, $\DL,\DT\sim|\ln(\beta
r)|$, but with $\beta\sim\lambda^{-1}\gg\kappa$ (dashed lines). At
large distances ($r\gg\lambda$) we have again $\DL,\DT\sim\pm 1/r^2$
(dotted curves) as a result of momentum loss to the immobile
inclusions.

Our predictions concerning the effect of immobile inclusions on the
self- and coupling diffusion coefficients of membrane proteins may be
directly checked in particle-tracking experiments. The effective
velocity Green's function derived above can be utilized
\cite{Camley2010} in other theories involving anchored membranes.  We
have found a strong effect of immobile inclusions already at a very
low area fraction (below one percent).  The effect is manifest in the
hydrodynamic screening length (Fig.\ \ref{fig_beta}B), the
self-diffusion coefficient of a mobile inclusion (Fig.\ \ref{fig_Ds}),
and the dynamic couplings between two mobile inclusions (Fig.\ 
\ref{fig_DLT} A vs.\ B). The high sensitivity to the presence of
immobile inclusions arises from the quasi-2D nature of the membrane
and the resulting long-ranged flows.
To highlight this special property, let us compare the situation to
its 3D counterpart. Consider a 3D fluid containing a volume fraction
$\phi$ of immobile particles of radius $a$. These obstacles introduce
a momentum-decay length, $\lambda\sim a\phi^{-1/2}$, which will
strongly affect the self-diffusion coefficient of a mobile particle
only when $\lambda$ becomes comparable to the particle size\,---\,\ie
for an appreciable value of $\phi$. Unlike the 3D case, which is
characterized by a single (small) length scale, $a$, the quasi-2D
dynamics depends on $a$ and another, much larger length scale,
$\beta^{-1}$. The long-range (logarithmic) nature of 2D flows [Eq.\ 
(\ref{G2D})] makes this length sensitive to the immobile inclusions
already for $\phi\sim 10^{-3}$ (Fig.\ \ref{fig_beta}A).  The quasi-2D
anomaly is reflected also in the nonanalytic decrease of $\beta^{-1}$
at small area fraction, as $\phi\ln\phi$.

The separation of length scales, $\beta^{-1}\gg a$, has also allowed
us to consistently include in the effective response of the membrane
all the monopolar hydrodynamic terms (\ie all orders of $\phi$) while
continuing to ignore spatial correlations in the distribution of
inclusions; for the low area fraction considered here ($\phi\lesssim
10^{-2}$) such spatial correlations are negligible. The other
assumption necessary to make this expansion valid is that the
distribution of inclusions is stationary, unaffected by the exerted
forces.
In addition, our analysis is restricted to the leading terms in
$\kappa a$, $a/r$, and $\beta a$. Corrections due to higher-order
terms in the first two factors are related to higher moments of the
force distribution exerted by individual inclusions. In actual
membranes $\kappa a$ is of order $10^{-3}$, and the approximation is
well justified. Higher-order terms in $a/r$ become arbitrarily small
at sufficiently large distances \cite{ft_deform}. By contrast, since
$\beta$ is density-dependent, the restriction $\beta a\ll 1$ limits
the validity of the analysis to $\phi\lesssim 10^{-2}$ almost
irrespective of $a$ (see Fig.\ \ref{fig_beta}A).

Immobile inclusions also affect the membrane's out-of-plane dynamics
\cite{Gov2003,Zhang2008}. To leading order in the normal deformations,
the in-plane and out-of-plane dynamics are decoupled
\cite{Levine2002}, and our quasi-2D model should remain valid within
the deformed surface. At higher orders of the deformation, or because
of inclusion--membrane curvature coupling, various corrections are
expected \cite{NajiBrown,ReisterSeifert,Dean2011}.

We should mention two additional ingredients of actual biomembranes.
The first is the filamentous network to which the immobile inclusions
are attached. Since the network is deformable, the anchoring
inclusions should not be completely immobile. Qualitatively, this
restricted freedom to move laterally will affect the membrane's
in-plane dynamics as if the immobile inclusions had a somewhat larger
effective size. In addition, the network adds a viscoelastic response,
which is important for frequency-dependent properties but vanishes for
the steady-state ones considered here.  It also introduces another
length scale\,---\,the network correlation length $\ell$\,---\,which
is of order $0.1$ $\mu$m \cite{biology}. This large length scale will
not significantly affect our results concerning the self-diffusion
coefficient, but will influence the coupling coefficients at $r>\ell$.
At such interparticle distances the immobile network can be replaced
by an effective substrate \cite{ourPRE}. The other ingredient is a
finite density of mobile inclusions in addition to the immobile ones.
Their effect can be readily incorporated by a proper renormalization
of $\etam$ and $\kappa$ \cite{ourBJ}.

We have seen that for a reasonable value of $\phi>(\kappa
a)^2|\ln(\kappa a)|\sim 10^{-5}$, the hydrodynamic interaction between
two mobile inclusions crosses over from a strong coupling [decaying
only as $\ln(1/r)$] for $r<\beta^{-1}$, to a suppressed coupling
(decaying as $1/r^2$) for $r>\beta^{-1}$ (Fig.\ \ref{fig_DLT}B). We
end with the following question. If the immobile inclusions were to
divide the membrane into effective domains, such that within a domain
two mobile inclusions are strongly coupled, whereas across domains the
coupling is suppressed, what would be the value of $\phi$?  Solving
the equation $a\phi^{-1/2}=\beta^{-1}$, we find $\phi\simeq 2\times
10^{-3}$ and $5\times 10^{-5}$ for $\kappa a=10^{-2}$ and $10^{-3}$,
respectively. For $a=5$ nm this corresponds to domain sizes of about
$0.1$ and $0.7$ $\mu$m. These large values stem again from the
quasi-2D nature of fluid membranes.

\begin{acknowledgments}
  We are grateful to Michael Kozlov for helpful discussions.  This
  research has been supported by the Israel Science Foundation
  (Grants Nos.\ 588/06 and 8/10).
\end{acknowledgments}




\begin{thebibliography}{99}
\bibitem{ft_shape} The results are logarithmically insensitive to the
  inclusion shape, so long as it is compact. In addition,
  due to the large viscosity contrast between the membrane and the
  outer fluid, only the part of the inclusion embedded inside the
  membrane is important.
\end{thebibliography}
\end{document}